# Corrected values of turbulence generated by general geothermal convection in deep Mediterranean waters

Hans van Haren

NIOZ Royal Netherlands Institute for Sea Research, P.O. Box 59, 1790 AB  Den Burg, the Netherlands.

hans.van.haren@nioz.nl

**Abstract**

A correction by a reduction factor O(100) is proposed for previously calculated turbulence values on unresolved convection-overturns induced by general geothermal heating in the deep Western Mediterranean. The correction includes modified application of reordering method for calculating turbulence values in convection turbulence with and without stratification above or below. The result is improved correspondence between geophysical determined heat flow through the seafloor and turbulence kinetic energy dissipation rate determined from high-resolution temperature sensors moored over 109 m in the overlying waters, with an average mixing coefficient of 0.5.

**1 Introduction**

The direct observation using moored high-resolution temperature (T-)sensors of buoyancy-driven convection turbulence attributed to general geothermal heating in the deep Western Mediterranean (van Haren 2023) unfortunately contains erroneous calculation of turbulence values like dissipation rate. The calculation uses the method of reordering instabilities to stable vertical density profiles (Thorpe 1977). The error in van Haren (2023) comprises a mix-up of heat- and buoyancy-flux, an over-complicating comparison between vertically integrated dissipation rate with geothermal heat flux and an incorrect transfer rate between density and temperature variations.

In this technical note, corrected calculations are presented that demonstrate turbulence values matching average geothermal heating well to within error. An observed difference by one order of magnitude in dissipation rate calculated for the case of geothermally induced unstable overturning exceeding the 109-m range of T-sensors and for that overlying by stable stratification is explained, using simple modelling, by the over-estimating effect of reordering for the former case. An effective correction is proposed for such unresolved overturns exceeding the T-sensor range.

**2 Moored T-sensor data**

As detailed in van Haren (2023), a T-sensor mooring was deployed at 42° 47′N, 06° 09′E, seafloor at z = -2480 m, about 40 km south of Toulon, France. The average local seafloor slope is less than 1º, a relatively flat topography 12 km seaward of the steep continental slope. T-sensors provided useful data through the winter of 2017/2018 when deep dense-water formation was not noticed in the mooring area. For calibration purposes and to establish the local temperature-density relationship, shipborne Conductivity-Temperature-Depth (CTD) profiles were obtained near the mooring site during deployment and recovery cruises.

The local deep Mediterranean waters are very weakly stratified, with buoyancy frequency N ~ 1f, f denoting the inertial frequency, so that typical temperature variations are 0.0001-0.001°C. Thus, T-sensor data require high precision and stability, through elaborate post-processing. As described in van Haren (2022) the common post-processing of calibration, long-term drift-correction and reference to nearby shipborne CTD-data are supplemented by reference to data from a period of near-homogeneous, temperature variability <0.0003°C, waters over the entire mooring range. Such near-homogeneity is found here on day 439. The data discussed in this note are from near that day (Fig. 1), to take most advantage of this reference before instrumental drift, of typically 0.001°C mo$^{-1}$ after aging, becomes larger than the environmental signal. As in van Haren (2023), the data are low-pass filtered at 1000 cycles per day in frequency to remove noise. In addition, 'short-term drift', actually imperfect contact between sensor and environment through the glass wall, are reduced using low-pass (0.1 cycle per m) vertical filtering under near-homogeneous conditions with variations O(0.0001°C) when geothermal heating is observable in overlying waters.

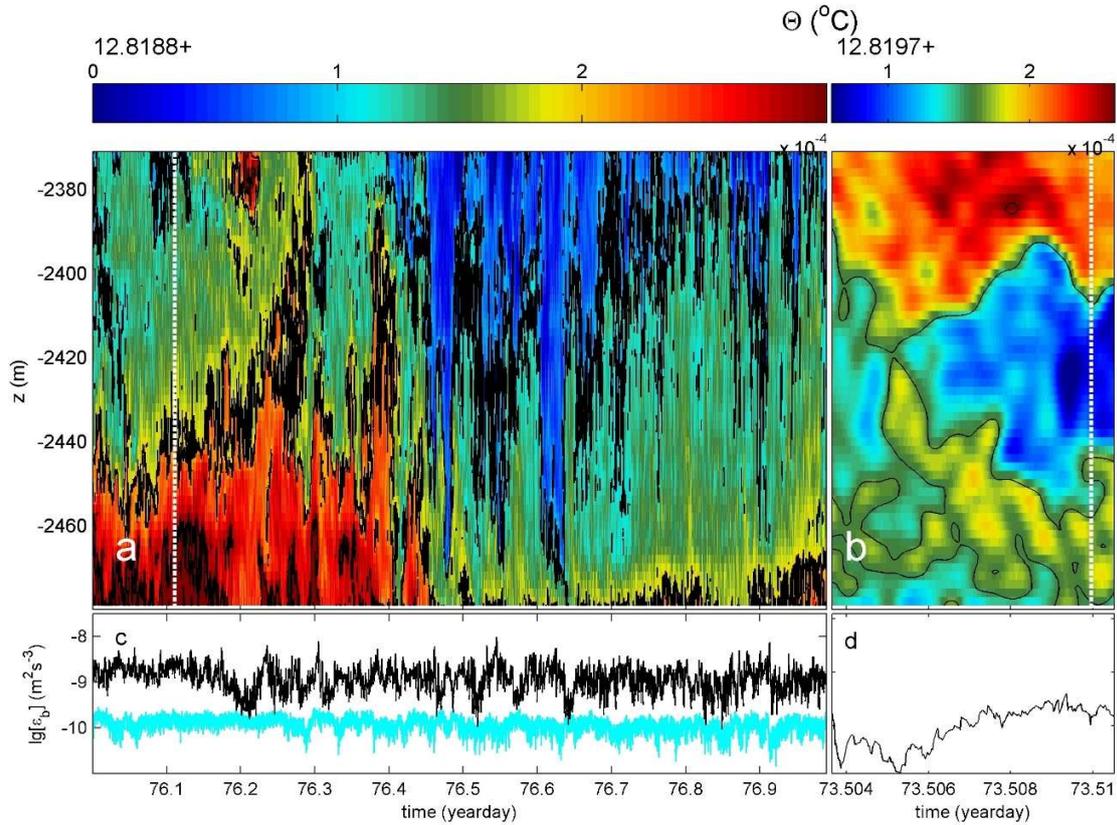

**Fig. 1** Short periods of convection turbulence of varying intensity and varying duration observed in the deep Western Mediterranean. (a) Replot of Fig. 6d in van Haren (2023) with slightly different colour and contour ranges. One day of the 109-m-vertical Conservative Temperature sampled at 2.0-m intervals, lpf at 1000 cpd and 0.1 cpm. The seafloor is at the horizontal axis. Black contours and value increments are every 0.0001 °C. The white dashed line indicates a single profile displayed in Fig. 3. (b) As a., but for 900 s of stratification over seafloor convection. (c) Logarithm of vertically averaged turbulence kinetic energy dissipation rate (black) and, in cyan after correction retaining per profile only values below threshold of $6.6\times10^{-10}$ $m^2s^{-3}$ of Fig. 4. (d) As c., but for data in b., without correction.

## 3 Results

**3.0 Observing the problem**

During about 25% of the 4.5 months record, moored T-sensor data showed such weakly stratified waters over the deep seafloor that convection turbulence associated with general geothermal heating was apparent in time-depth plots (van Haren 2023). At the particular mooring site during the particular autumn/winter, the convection turbulence often exceeded the h = 109 m range of T-sensors above the seafloor albeit vertical plumes are generally broken and slanted (Fig. 1a). Although convection plumes are accompanied by secondary shear (Li and Li 2006), average spectra showed significant deviation from the inertial subrange of shear-induced turbulence with a dominance of the buoyancy subrange of convection turbulence throughout the 109-m vertical range (van Haren 2023). It is noted that indeed all dynamics occurred within a temperature range of only a few 0.0001°C.

Occasionally, the convection turbulence was underneath well-stratified waters (depressed) within the range of T-sensors (Fig. 1b). Whilst such a condition of stratification overlying deep convection commonly occurs, except perhaps during rare brief moments of deep water formation in exceptional late winters, and considering that general geothermal heating has the same value when averaged over sufficient time, it is expected that the two data periods of Fig. 1 provide the same value of convection turbulence induced above the seafloor. However, applying the method by Thorpe (1977) of reordering unstable density (temperature) profiles to calculate turbulence values (Appendix) results in considerably different, by one order of magnitude, time series of vertically averaged near-bottom 'b' turbulence kinetic energy dissipation rate [$\varepsilon_b$] equals in this case [$\varepsilon_h$] = $h^{-1}\int\varepsilon dh$ (compare the black graphs in Fig. 1c,d). Here, the vertical averaging is covering the largest convective overturn when stable stratification is above the T-sensors, which is a prerequisite for quantitative use of Thorpe (1977)'s method. Near the sea surface, overturn scales are found useful to determine the depth of nighttime convective mixing (Brainerd and Gregg 1995; Kumar et al. 2021). As a physical cause of stratification blocking the turbulence is unlikely explaining the order of magnitude difference, a technical cause of overestimating turbulence by unresolved overturns under changing preconditioning is anticipated, and investigated below.

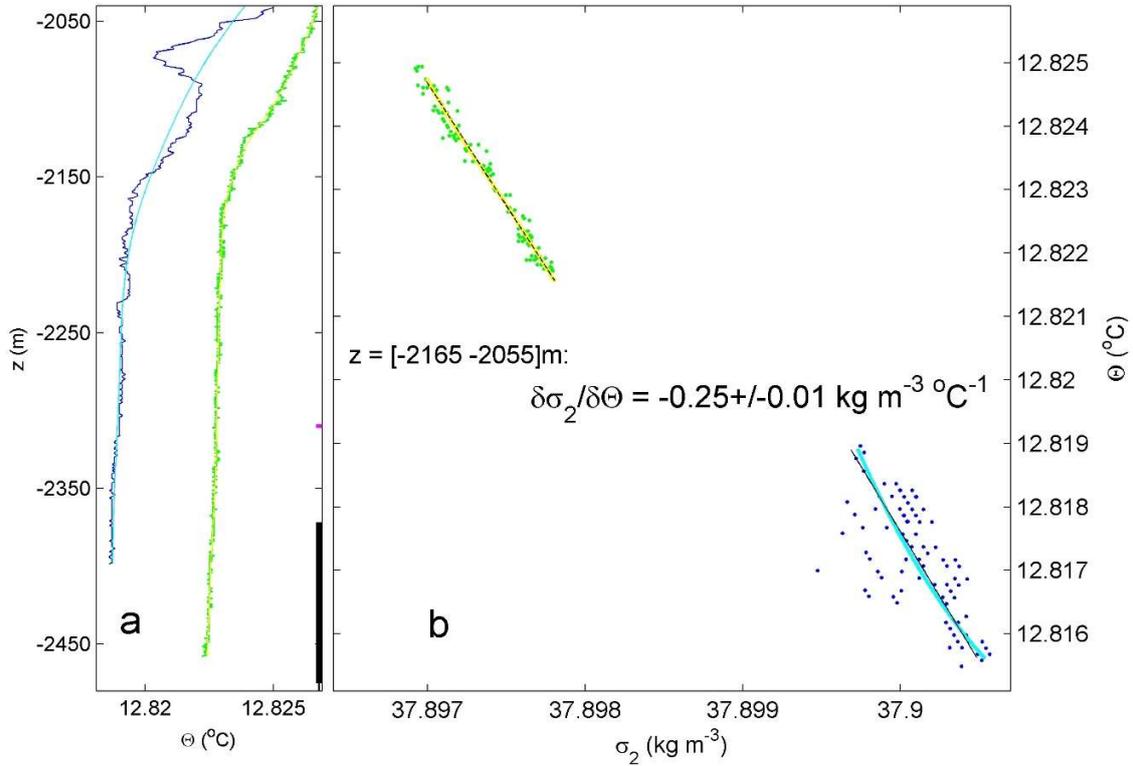

**Fig. 2** Local deep Mediterranean CTD observations. (a) Conservative Temperature from about lower 400 m above the seafloor from two different years, with the black bar denoting the height of the mooring. (b) Plot of relationship between Conservative Temperature and density anomaly referenced to pressure level of $2\times10^7$ Pa from shipborne CTD observations in 2018 (blue) and 2020 (green). Average slopes are determined in weakly stratified waters where buoyancy frequency N = 2-2.5f, f the inertial frequency, over the indicated range about 350 m above the seafloor. Smooth data lines are 50-m lpf versions of the 1-m averaged data (dots). In black are the straight best-fit slopes to the smooth data.

**3.1 CTD-determined temperature-density consistency**

In the lower 400 m above the seafloor, the shipborne CTD-profiles vary, not only on small 10-m vertical scales, but also on large 100-m vertical scales (van Haren 2023). Although commonly a local relationship is established between Conservative Temperature (IOC et al. 2010) $\Theta$ and density anomaly $\sigma_2$ referenced to a pressure level of $2\times10^7$ N m$^{-2}$ around the moored T-sensor range, this results in (too) large, >10%, errors in waters where N ~ 1f, such as occasionally found near the mooring site (Fig. 2a). Thus, one is obliged to establish a relationship from layers higher up where N = (2-3)f, and which regularly reach the moored T-sensors, from above or from the sides.

Smoothed over a 110-m vertical range between -2165 < z < -2055 m the $\Theta$-$\sigma_2$ consistent relationship amounts (Fig. 2b),

$$\delta\sigma_2/\delta\Theta = -0.25\pm0.01 \text{ kg m}^{-3} \text{ °C}^{-1}. \tag{1}$$

The relatively tight relationship (1) is a factor of 3.4 smaller than used in van Haren (2023) which results in calculated (Appendix) turbulence dissipation rate values reduced by a factor of 6.3.

### 3.2 Geothermal buoyancy flux

The average local general geothermal heat flux amounts Q = 0.11±0.03 W m$^{-2}$ (Pasquale et al. 2016; Ferron et al. 2017). This is the heat flux transferred from the sediment into the overlying waters, presumably via conduction. For simplicity, it is considered a constant flux without variations in time or space, ignoring complex geological and geophysical processes (e.g., Kunath et al. 2021). It is expected that this heating from below will start convection turbulence in the overlying waters when the stable stratification supply from above is sufficiently weak, i.e. (pre)conditioning is near-neutral. We assume no other governing turbulence processes during such periods.

As the turbulence is buoyancy driven, the heat flux is transferred to buoyancy flux,

$$J_b = g\alpha Q/\rho c_p = 6\times 10^{-11} \text{ m}^2\text{s}^{-3}, \tag{2}$$

where g denotes the acceleration of gravity, $\alpha$ = 2.3×10$^{-4}$ °C$^{-1}$ the local thermal expansion coefficient, $\rho$ the density of overlying water and $c_p$ = 3950 J kg$^{-1}$ °C$^{-1}$ the local heat capacity.

For nighttime near-surface convection, Lombardo and Gregg (1989) established a relationship between 'mixing layer' turbulence dissipation rate $\varepsilon_m$ and positive buoyancy flux cooling $\varepsilon_m$ = (0.5±0.1)$J_{b,s}$. This ratio excluded the upper 'surface layer' (s) 20 m showing one to two orders of magnitude larger $\varepsilon_s = (10\text{-}100)\varepsilon_m$ that included wave action besides convection. Since we do not have microstructure profiler data from near the deep seafloor, the overturn-averaging (Appendix) will include data from the entire relevant distance above the seafloor starting at h = 1 m.

Thus, assuming a continuous buoyancy flux transferred from the solid seafloor into overlying waters we expect a relationship of,

$$J_b = \Gamma_e[\varepsilon_b], \tag{3}$$

where $\Gamma_e = 0.2\text{-}1.0$ represents the mixing coefficient with values between 0.2 for shear turbulence in the ocean (Gregg 2018), via 0.3-0.7 for convection induced by breaking internal waves over sloping topography (Alford et al. 2024), and 1.0, yielding a mixing efficiency of 0.5, for pure convection as in Rayleigh-Bénard convection and Rayleigh-Taylor instabilities (Dalziel et al. 2008; Gayen et al. 2013; Ng et al. 2016). Hence, an expected value for geothermal induced turbulence dissipation rate $[\varepsilon_b] = \varepsilon_{GH}$ is calculated from (2) in (3) of,

$$\varepsilon_{GH} = J_b/\Gamma_e = 0.6\text{-}3.0\times10^{-10} \text{ m}^2\text{s}^{-3}. \tag{4}$$

For comparison, the values in (4) are three orders of magnitude lower than the outgoing radiative buoyancy flux near the sea surface.

**3.3 Observing complete and incomplete overturns of convection turbulence**

The time-mean turbulence dissipation rate value averaged over the lower 55 m above the seafloor from observations in Fig. 1b amounts,

$$[\varepsilon_b] = 1.2\pm0.7\times10^{-10} \text{ m}^2\text{s}^{-3}, \quad \Gamma_e = 0.5\pm0.3. \tag{5}$$

This value is well within error corresponding with that of geophysical determined geothermal heating $\varepsilon_{GH}$ in (4), and suggesting a mixing efficiency commensurate dominant convection turbulence.

A representative vertical temperature profile (Fig. 3a) shows general instability up to half the T-sensor range that results in ragged short vertical scale variations in displacements transferred in terms of dissipation rate values (Fig. 3b). Because of the overlying stable stratification, the largest overturn of the lower-half instability is 'complete' and well resolved. Note that these individual dissipation rate values have no quantitative meaning, as averages should be taken over entire overturns, in the vertical or in time. However, the shape of the profile may be considered qualitatively. It is found rather homogeneous in the vertical, despite the short-scale variations. It is not very different from the smoother 900-s time-mean profile.

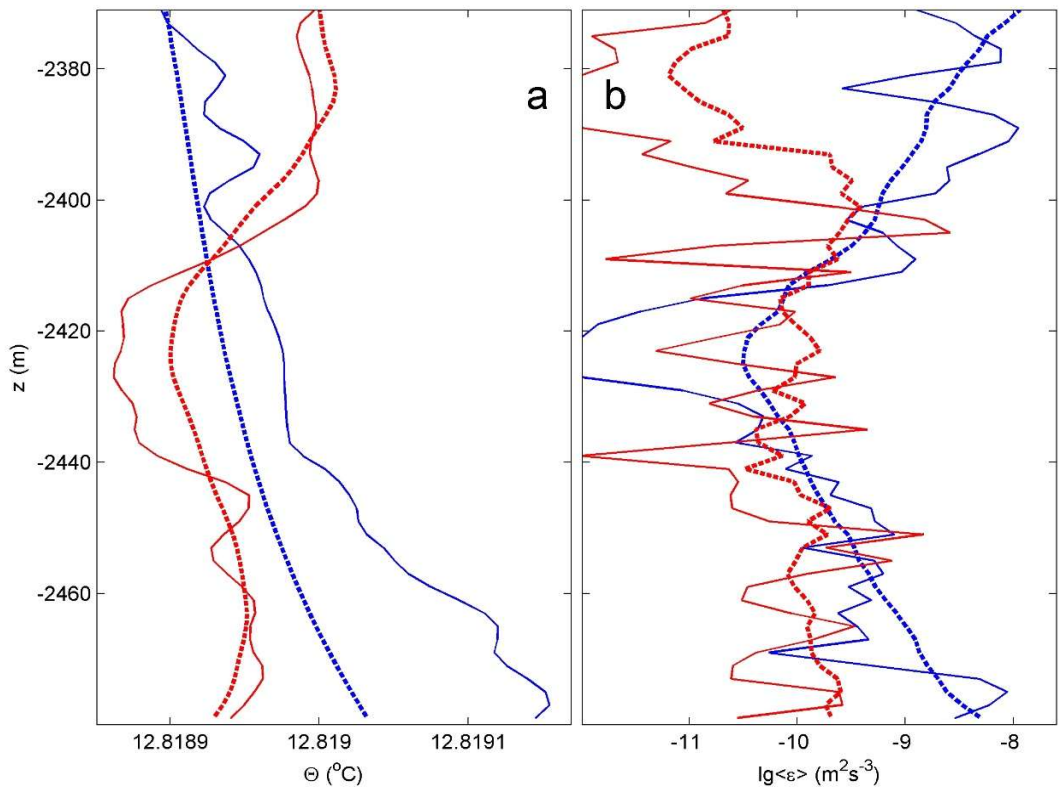

**Fig. 3** Time-mean (thick-dashed) and single (thin-solid) profiles for data in Fig. 1a,b (blue) and Fig. 1c,d (red). (a) Conservative Temperature. (b) Turbulence kinetic energy dissipation rate.

In contrast, a representative single profile from the data in Fig. 1a shows general instability over the entire 109-m T-sensor range, albeit with similar rate over most of its lower half as the profile from data in Fig. 1b. However, the associated dissipation rate profile is distinctly different from that of Fig. 1b data, with less small-scale variations, a large-scale minimum in the center and large-scale maxima at the upper and lower edges for the data in Fig. 1a. The single profile is part of a less extreme and smoother 1-day mean profile that retains the center-minimum and edge-maxima. Obviously, the largest overturns are 'incomplete' and not well resolved as they may exceed the T-sensor range. The profiles of displacement-determined dissipation rates qualify as artificial.

**3.4 A simple model of (in)complete convection overturns**

Simple modeling was performed for better understanding the effects of calculation of turbulence values from vertical density profiles using Thorpe (1977) reordering of unstable layers that are unbounded by stratification, such as in convection turbulence. An 'unstable' profile model of incomplete overturn is compared with two 'stratified' profile models of different levels of stratification overlying the unstable layer. The models represent geothermal heating observations of convection turbulence under different overlying stratification preconditioning in Fig. 1, 3, but they can also be used for similar convection turbulence from above during (nighttime) cooling near the surface.

The 1D-models use 2-m vertical distancing over 55 data points between h = 1-109 m above the seafloor, like a single profile in the moored T-sensor observations (Fig. 4). The models follow the same temperature gradient as in observed single profiles of Fig. 3a, and are converted to density profiles using the density-temperature relationship (1) observed in immediate overlying stratification. The blue unstable model has a monotonic temperature decrease with height. The green (g) stratified model just compensates the instability below, while the magenta (m) stratification is twice as large and corresponds with that of the observed stable profile in Fig. 3a. As a result, it is expected that the two stratified models g,m result in turbulence dissipation rate of $\varepsilon_{modg,m} \approx 1.2\times10^{-10}$ m$^2$s$^{-3}$ as in (5).

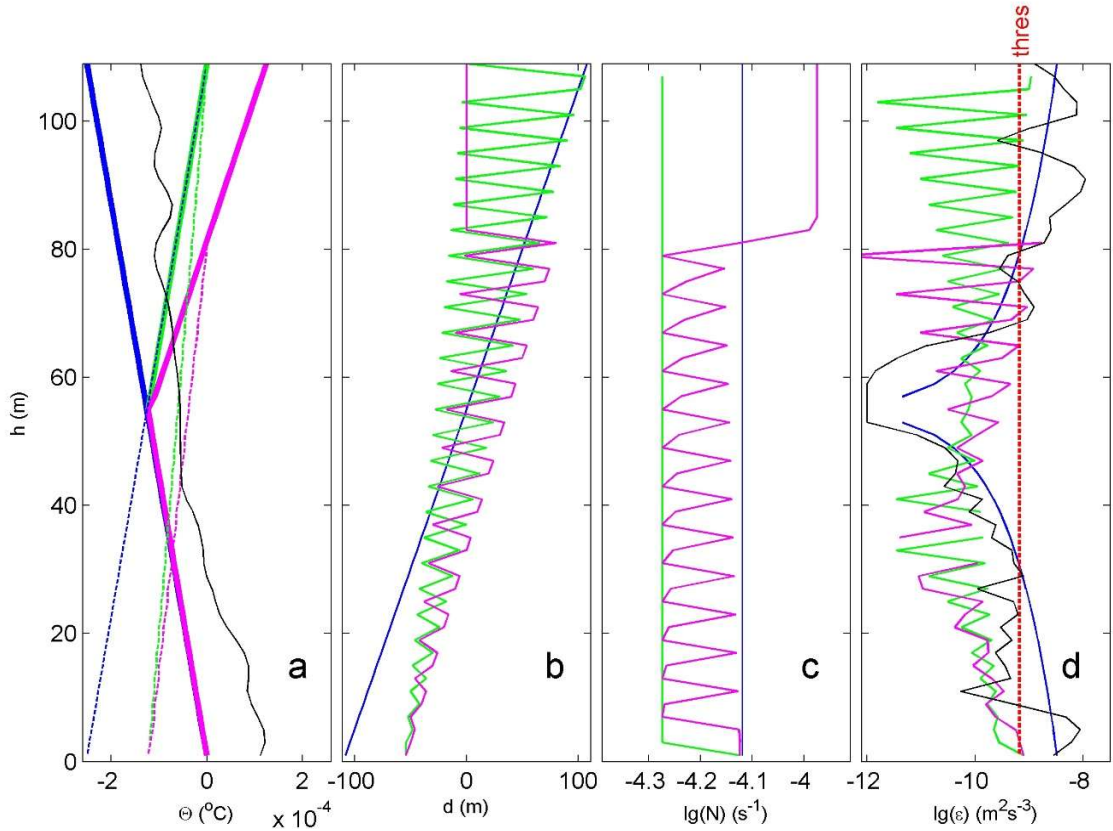

**Fig. 4** Model profiles to investigate overturning effects with and without stratification over-capping an unstable layer above the seafloor, here at height h = 0. (a) Conservative Temperature anomaly for three models: unstable over the entire range (blue), unstable at the same rate over bottom-half overlying with stable stratification at the same rate (green), unstable at the same rate over bottom-half overlying with double-rated stratification (magenta). The dashed curves are their stably reordered profiles. In solid-black is the observed single profile from unstable data in Fig. 1a. (b) Displacements after reordering the model profiles of a. (c) Logarithm of buoyancy frequency from the reordered model profiles in a. (d) Logarithm of calculated dissipation rate profiles for the profiles in a. The mean values below threshold for the unstable blue and black profiles are similar to within error to the mean of values below threshold (red-dashed) of the lower half of the green and magenta profiles. The threshold equals the vertical means ratio of blue and green or magenta dissipation values times the observed mean value of Fig. 1d (see text).

All the dynamics are comprised in a vertical relative temperature variation of $\pm 2\times 10^{-4}$ °C (Fig. 4a). Reordering of the model profiles demonstrates that none of the profiles becomes neutral, all are weakly, but non-negligibly stratified. From the displacements (Fig. 4b) we learn that the fully unstable model yields a monotonic profile describing one large symmetric-around-zero overturn albeit of unresolved extent, while both stratified models flip back and forth in displacement value over short vertical distances describing an asymmetric fully resolved large overturn.

Technically, the inference is that, for the green stratified model, one displaced particle is adjacent to a non-reordered one. (Physically we miss the unknown erosion of stratification in the unstable model). The rms displacement values between unstable blue model and stratified green model differ by about a factor of 0.7 (Fig. 4b). The rms buoyancy frequency also differs by a factor of 0.7 between the unstable and the green stratified models, while being similar to the unstable value for the magenta stratified model (Fig. 4c).

The differences between the unstable and stratified models yield a theoretical difference in turbulent kinetic energy dissipation rate by a factor of $0.7^5 = 1/5.6$ following (A1). Due to finite sampling and edges the practical factor between the models amounts about 1/5.5.

The profiles of turbulence dissipation rate (Fig. 4d) show a vertically symmetric and smooth shape for the unstable model with low values in the middle and high values at the edges. The stratified models result in vertically more distributed dissipation values, albeit with shorter scale variations resulting in less smooth profiles. These two characteristic differences between unstable and stratified models is also seen in observed single profiles from the two periods in Fig. 3b, of which the unstable profile is replotted in Fig. 4d. Since there is no expected difference in (long-term) average geothermal heat flux at a particular location affecting turbulence in waters with or without overlying stratification across the range of convection, a correction is sought for dissipation rate values of the unstable model.

The applied correction removes values larger than a proposed threshold of $5.5 \times \varepsilon_{GH}$ from the mean, where we use dissipation rate by geothermal heat flux (5). This corresponds to a mean turbulence dissipation rate from displacement values between approximately $30 < h < 75$ m in Fig. 4d, for the unstable profiles of both the model and the single observed profile. Applying this correction to the data calculated from Fig. 1a gives the cyan time series of mean dissipation rate values in Fig. 1b. Its time mean value of $1.25\times10^{-10}$ $m^2s^{-3}$ is a reduction by a factor of 11 to the mean of black time series in Fig. 1a. It is close to that of Fig. 1d of $1.23\times10^{-10}$ $m^2s^{-3}$, well within error bounds of $0.7\times10^{-10}$ $m^2s^{-3}$.

The above implies that unresolved overturns by too short profiling yield over- rather than under-estimates (van Haren, 2023) of turbulence dissipation rate values. The over-estimates can be corrected using the proposed threshold. The comparison with geophysical determined geothermal heat flux confirms successful application of reordering method (Thorpe 1977) to establish convection turbulence values in deep-sea waters, as has been demonstrated for near-surface waters (Kumar et al. 2021).

**Acknowledgments** I thank captain and crew of R/V l'Atalante and NIOZ-NMF for their assistance during deployment and recovery and for the construction of the mooring array. I thank M. Stastna (Univ. Waterloo, Canada) for providing the 'darkjet' colour-map suited for T-sensor data.

**Funding** This research was supported in part by NWO, the Netherlands Organization for the advancement of science.

**Data availability** Data that support the findings of this study are available from the corresponding author, upon reasonable request.

**Conflict of interest** The author declares no competing interests.

**Appendix. Moored T-sensor turbulence values**

The relatively tight relationship (1) implies the moored T-sensor data may be used as a proxy for density variations and in which salinity contributions are implicitly incorporated. The relationship is used to calculate turbulence values following the method of reordering unstable data-points to monotonously stable vertical profiles (Thorpe, 1977). Turbulent overturns follow reordering every 2 s the 109-m high potential density profile $\sigma_2(z)$ into a stable monotonic profile $\sigma_2(z_s)$. Displacements $d = \min(|z-z_s|)\cdot\text{sgn}(z-z_s)$ are calculated necessary for generating the reordered stable profile. The turbulence kinetic energy dissipation rate reads,

$$\varepsilon = 0.64 d^2 N^3, \tag{A1}$$

where buoyancy frequency $N$ is computed from each of the reordered, essentially statically stable, vertical density profiles.

The numerical constant follows from empirically relating the root-mean-square (rms) overturning scale $d_{rms} = (\Sigma d^2/n)^{0.5}$ over n samples with rms-Ozmidov scale

$$L_O = (\varepsilon/N^3)_{rms}$$

of largest isotropic turbulence overturns in a stratified fluid as an average over many realizations via the ratio: $L_O/d_{rms} = 0.8$ (Dillon, 1982). This ratio reflects turbulence in any high Reynolds number stably stratified environment like the deep-sea, in which shear-driven and convection-turbulence intermingle at small and large scales and are difficult to separate. In all cases, the mechanical turbulence must work against the stratification that follows from the reordering. It has thus successfully been applied for mainly convection-turbulence (e.g., Chalamalla and Sarkar 2015; Kumar et al. 2021) while first used for mainly shear-turbulence (Thorpe 1977). The method works for convection-turbulence, provided sufficient stratification exists in the observed profile above and/or below, for complete reordering. As outlined in this note, incompletely reordered unstable overturns yield overestimates of turbulence values, for example in moored instrumentation of limited vertical range.

In (A1), individual d are used rather than taking their rms-value across a single overturn as originally proposed by Thorpe (1977). The reason is that individual overturns cannot easily be distinguished, first, because they are found at various scales with small ones overprinting larger overturns, and second, because some overturns exceed the range of T-sensors. For quantification of turbulence values, 'sufficient' averaging is required, also to include various turbulence types of different scales and different age with potentially different $L_O/d_{rms}$-ratio (Chalamalla and Sarkar, 2015) during a turbulent overturn lifetime. While shipborne vertical profiling instruments limit to vertical data averaging, indicated by [.], the advantage of a densely instrumented mooring line is also averaging data over time, indicated by <.>.

Kunath P, Chi W-C, Berndt C, Liu C-S (2021) A rapid numerical method to constrain 2D focused fluid flow rates along convergent margins using dense BSR-based temperature field data. J. Geophys. Res. Solid Earth 126:e2021JB021668.

Li S, Li H (2006) Parallel AMR code for compressible MHD and HD equations. T-7, MS B284, Theoretical division, Los Alamos National Laboratory. https://citeseerx.ist.psu.edu/pdf/03e1663486594ce991cc4bbdffa031dbbeb3ab33. Accessed 14 December 2024

Lombardo CP, Gregg MC (1989) Similarity scaling of viscous and thermal dissipation in a convecting surface boundary layer. J Geophys Res 94:6273-6284

Ng CS, Ooi A, Chung D (2016) Potential energy in vertical natural convection. Proc 20th Australas Fl Mech Conf, 727 (pp 1-4)

Pasquale V, Verdoya M, Chiozzi P (1996) Heat flux and timing of the drifting stage in the Ligurian–Provençal basin (northwestern Mediterranean). J Geodyn 21:205-222

Thorpe SA (1977) Turbulence and mixing in a Scottish loch. Phil Trans Roy Soc Lond A 286:125-181

van Haren H (2022) Thermistor string corrections in data from very weakly stratified deep-ocean waters. Deep-Sea Res I 189:103870

van Haren H (2023). Direct observations of general geothermal convection in deep Mediterranean waters. Ocean Dyn 73:807-825.
17